# Exploring impulsive solar magnetic energy release and particle acceleration with focused hard X-ray imaging spectroscopy

**Steven Christe**[1], S. Krucker[2,3], L. Glesener[4], A. Shih[1], P. Saint-Hilaire[2], A. Caspi[5], J. Allred[1], M. Battaglia[3], B. Chen[6], J. Drake[7], B. Dennis[1], D. Gary[6], S. Gburek[8], K. Goetz[4], B. Grefenstette[9], M. Gubarev[10], I. Hannah[11], G. Holman[1], H. Hudson[2], A. Inglis[1], J. Ireland[1], S. Ishikawa[12], J. Klimchuk[1], E. Kontar[11], A. Kowalski[13], D. Longcope[14], A. Massone[15], S. Musset[4], M. Piana[15], B. Ramsey[10], D. Ryan[1], R. Schwartz[1], M. Steślicki[8], P. Turin[2], A. Warmuth[16], C. Wilson-Hodge[10], S. White[17], A. Veronig[18], N. Vilmer[19], T. Woods[20]

[1] NASA Goddard Space Flight Center [2] Space Sciences Lab, Univ. of California at Berkeley [3] Univ. of Applied Sciences NW Switzerland [4] Univ. of Minnesota [5] Southwest Research Institute [6] New Jersey Institute of Technology [7] Univ. of Maryland [8] Space Research Centre of the Polish Academy of Sciences [9] Caltech [10] NASA Marshall Space Flight Center [11] Univ. of Glasgow [12] Institute of Space and Astronautical Science (ISAS)/JAXA [13] Colorado University/NSO [14] Montana State University [15] Univ. of Genoa [16] Leibniz Institute for Astrophysics Potsdam [17] Air Force Research Lab [18] Univ. of Graz [19] Observatoire de Paris [20] LASP

How impulsive magnetic energy release leads to solar eruptions and how those eruptions are energized and evolve are vital unsolved problems in Heliophysics. The standard model for solar eruptions (**Figure 1**) summarizes our current understanding of these events. Magnetic energy in the corona is released through drastic restructuring of the magnetic field via reconnection. Electrons and ions are then accelerated by poorly understood processes. Theories include contracting loops, merging magnetic islands, stochastic acceleration, and turbulence at shocks, among others (e.g., **[1][2][3][4]**, see **[5]** for review). Some accelerated particles can escape into the heliosphere to be observed as solar energetic particle (SEP) events *in situ*, and can have marked space-weather impacts. Other particles travel downward to the footpoints of magnetic loops in the dense chromosphere, where they deposit most of their energy. This energy rapidly heats chromospheric material to tens of millions of degrees, which then "evaporates" up into the corona. As magnetic fields reconnect, a coronal mass ejection (CME) can be released which can contain >$10^{12}$ kg of material traveling at speeds often exceeding 1000 km s$^{-1}$ and can also accelerate charged particles via shock-driven processes. These events are known to be the fundamental cause of hazardous space weather, which can damage satellite assets, terrestrial power grids, and pose a hazard to human and robotic space explorers.

Although this basic model is well established, the fundamental physics is poorly understood. HXR observations using grazing-incidence focusing optics can now probe all of the key regions of the standard model (see black circles in **Figure 1**). These include two above-the-looptop (ALT) sources which bookend the reconnection region and are likely the sites of particle acceleration and direct heating. The flux rope or CME core is also known to be a site of accelerated particles and hot plasma. Transport mechanisms affect the particles as they travel away from the Sun or down the loop legs and are finally stopped by the dense chromosphere leading to hot flare loops. RHESSI has provided some of the most important observations and tests of the standard flare model **[6]**. This is because HXR imaging and spectroscopic observations provide direct and accurate information on the location and energetics of accelerated electrons since *bremsstrahlung* X-ray emission cross-sections are well known and X-rays pass unimpeded through the optically thin corona from their point of origin. Moreover, thermal X-ray line and continuum emissions are not significantly affected by non-equilibrium ionization, which can affect interpretations of the plasma temperature using other wavelengths.

However, the limited dynamic range and sensitivity afforded by indirect (Fourier-based) imaging used by RHESSI [7] and other instruments, such as the Hard X-ray Telescope (HXT) on Yohkoh [8], has not allowed key processes such as particle acceleration and direct heating to be studied concurrently. This is because X-ray emission from the corona near the source of energy release and acceleration is relatively faint and cannot be separated by these instruments from the associated, but much brighter, chromospheric footpoint sources. In the rare cases when RHESSI observed coronal sources, the footpoint sources were either unusually weak in comparison or were over the limb and hence not visible [9]. New HXR focusing optics can now measure the X-ray emission from electrons as they are accelerated in the corona and propagate along the magnetic field lines, at the same time as the much brighter emission from the chromospheric footpoints where electrons lose the majority of their energy. Consequently, for the first time, it is now possible to fully characterize the accelerated electrons and hottest plasmas as they evolve in energy, space, and time.

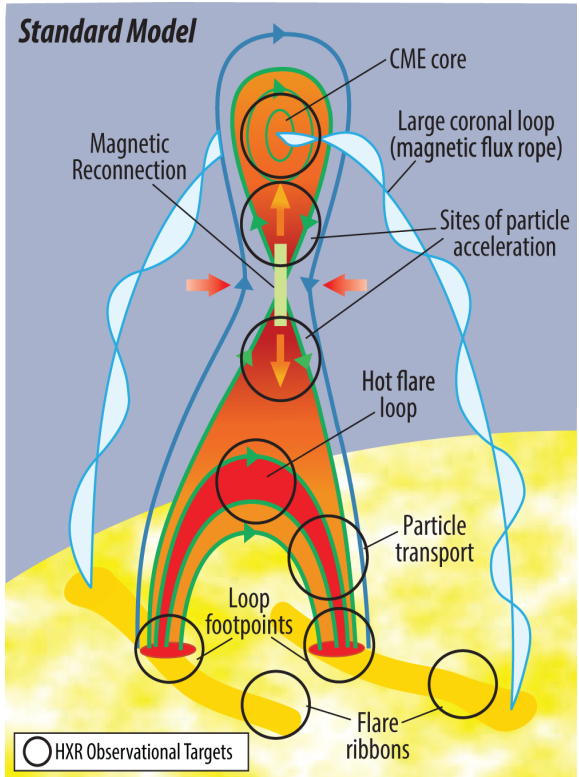

**Figure 1**. HXR observations emission from all key points in the standard model of solar eruptive events.

Direct imaging spectroscopy of thermal and non-thermal coronal and chromospheric X-ray sources simultaneously can enable the first systematic observations of non-thermal sources at and above the tops of flaring magnetic arcades even in the presence of much stronger X-ray footpoint sources. Some particle acceleration theories, e.g., those involving contracting and merging magnetic islands in the reconnection outflow [2][10], directly associate these sources with acceleration regions. They predict that particle acceleration can occur on timescales of 0.5–5 seconds [2][10]. Other theories predict acceleration sites in current sheets far above the arcades (e.g., [11]) or in the looptops. HXR focusing optics can easily image such rapid time variations.

The fraction of electrons accelerated out of the ambient Maxwellian velocity distribution is an essential constraint on acceleration models. For example, the merging magnetic island theory [2] or acceleration by super-Dreicer electric fields in a reconnecting current sheet [12] can accelerate a large fraction of the available electrons, whereas models that invoke acceleration by large-scale sub-Dreicer electric fields [13] accelerate only a small fraction of the electrons. The transition from thermal to non-thermal regions of the electron spectrum can be determined from the spectrum of X-rays emanating from the acceleration region. RHESSI confirmed the existence of these ALT sources but, because of its limited dynamic range, this was only possible in a few of the many thousands of flares it observed. Interpretation of the spectra of these coronal sources has suggested that all electrons in the acceleration region are accelerated in some events [14][15] in a bulk energization process.

Direct HXR imaging telescopes can measure HXRs from electrons as they are accelerated in the corona and propagate along magnetic field lines at the same time as the much brighter emission from chromospheric footpoints where the electrons lose most of their energy. With HXRs one can observe heated plasma and energetic electrons as close to the energy release site as possible, allowing key processes, such as particle acceleration and direct heating, to be accurately diagnosed. Furthermore, HXR spectra can characterize accelerated electrons along the paths traveled from the corona to the chromosphere, obtaining previously unavailable information about transport effects and associated heating.

The science achievable by a direct HXR imaging instrument can be summarized by the following science questions and objectives which are some of the most outstanding issues in solar physics

**(1) How are particles accelerated at the Sun?** (1a) Where are electrons accelerated and on what time scales? (1b) What fraction of electrons is accelerated out of the ambient medium?

**(2) How does magnetic energy release on the Sun lead to flares and eruptions?**

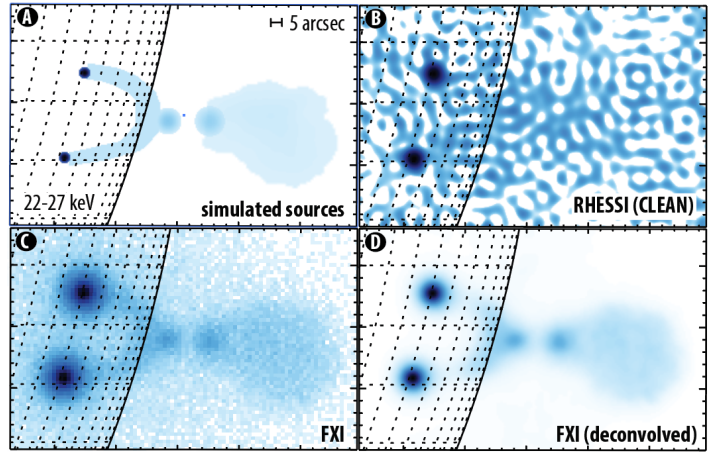

**Figure 2**. HXR focusing optics can easily observe faint signatures of energetic electrons directly in the corona as they are being accelerated. (A) The input sources for the simulation. The sizes and and intensities are based on observations. (B) Simulated RHESSI CLEAN image (C) Simulated image using HXR focusing optics shows all sources. (D) Deconvolution cleanly separates the sources and provides straightforward imaging spectroscopy.

The superior imaging provided by HXR focusing optics is illustrated in Figure 2. This simulation is based on the event shown in **Figure 1** combined with the capabilities of a recently proposed instrument (FOXSI/FXI) on a NASA Small Explorer mission. This instrument concept is the culmination of a series of successful projects and missions. High-resolution focusing optics combined with Japanese-developed strip detectors flew on the FOXSI-1 and -2 sounding rockets in 2012 **[16]** and 2014 **[17]**. This was preceded by earlier developments on five flights of the High Energy Replicated Optics (HERO) balloon payload **[18][19][20]**. New detectors recently developed at GSFC in collaboration with MSFC and RAL are also available which can handle high counting rate **[21]** and do not require any new development for space-based application (TRL-6).

The importance of these science goals are reflected by the Heliophysics Decadal Survey and associated NASA Heliophysics Roadmap goals. This instrument concept was recommended by the solar subpanel of the Heliophysics Decadal Survey as part of the SEE 2020 mission **[22]**. The report further stated that such an instrument would provide "tremendous new science." A Focusing Optics X-ray Solar Imager (FOXSI) instrument, which can be built now using proven technology and at modest cost, would enable revolutionary advancements in our understanding of impulsive magnetic energy release and particle acceleration, a process which is known to occur at the Sun but also throughout the Universe. Such an instrument would combine the strengths of organizations including JAXA that have collaborated effectively in the past.